\documentstyle[12pt]{article}
\begin{document}
\begin{titlepage}
\title{{\bf Quasilocal energy conditions}}
\author{Geoff Hayward\thanks{e-mail:
hayward@phys.ualberta.ca}\\
Institute of Theoretical Physics,\\
412 Avadh Bhatia Physics Laboratory,
University of Alberta,\\
Edmonton, AB,
Canada T6G2J1}
\date{PACS: 04.20.Cv, 04.20.Fy, 95.30.Sf}
\maketitle
\begin{abstract}
The classical value of the Hamiltonian for a system
with
timelike boundary has been interpreted as a
quasilocal energy.
This quasilocal energy is not positive definite.
However, we derive a `quasilocal
dominant energy condition'
which is the natural consequence of the
local dominant energy condition.
We discuss some implications of this quasilocal energy condition.
In particular, we
find that it implies a `quasilocal weak energy condition'.
\end{abstract}
\end{titlepage}
\setcounter{page}{1}
\baselineskip=10.mm
\section{Introduction\label{S1}}
\pagestyle{plain}
While the concept of energy plays a fundamental role in
the classical dynamics of finite systems, it has
not been clear how to extend this concept into
the setting of general relativity.
Traditionally, the relativist has had to choose
between a `global' energy---the Arnowitt--Deser--Misner
(ADM) mass defined at spacelike infinity---and a
`local' energy---the energy density defined
over an infinitesimal neighbourhood.  The search
for a `quasilocal' definition of energy which
bridges the gulf between these extremes has
uncovered a number of different candidates\footnote{
An excellent list of references which samples the
literature on local and quasilocal energy is given in the
first citation of Ref.~\cite{Brown/York}.}.
However, the issue of whether one of these
has preferred status over the others
has yet to be settled.

Recently, Brown and York~\cite{Brown/York} have proposed that the
quasilocal energy for a system
(as measured by a given set of observers)
be defined as the classical value of its Hamiltonian.
This definition, generalizes
the definition of energy which appears in mechanics
and
has a number of attractive features.
Many of these have been outlined by Brown and
York.

Here our purpose is not to review all the various features
of this definition of quasilocal energy, nor to judge its
merits by comparison to other definitions.
Quite apart from any interpretation as a quasilocal energy,
the Hamiltonian plays a fundamental role in general
relativity.
It is natural to
wonder what consequences the local energy conditions
hold for the
classical value of the Hamiltonian.
Placing labels aside, this is the question we address.

We demonstrate that when the local dominant
energy
condition holds everywhere on a spacelike hypersurface,
a
`quasilocal dominant energy condition' can be derived.
Defining a `ground state' as
a state which extremizes
the Hamiltonian over the class of vacuum states
which all have a given boundary geometry,
the quasilocal dominant energy condition
implies that perturbations from
a ground state which leave
boundary conditions invariant cannot give rise to
spacelike variations
of the quasilocal momentum field.
A consequence of this is that such perturbations
cannot decrease
the quasilocal energy below its ground state value.

While the details of this quasilocal energy condition
will be specified below, it is
important to stress at the outset
that it does not imply positivity
of the quasilocal energy.
In fact, it is easy to show that
the quasilocal energy is not positive definite.  Explicit
examples of negative energy states can be found in
Ref.~\cite{Euclidean}.

To see why the quasilocal energy cannot be positive
definite, consider a system of metric and matter fields
with support on a compact spacelike hypersurface without
boundary.  The extremal Hamiltonian on such a hypersurface
vanishes by virtue of the constraint equations.
Now let a closed 2--surface bifurcate the spacelike hypersurface
into `inside' and `outside'
regions.
The Hamiltonians for the
`inside' and `outside' regions will in general be non--vanishing
(by virtue of boundary contributions to the Hamiltonian).
Furthermore, since the Hamiltonian is additive,
the values for the `inside' and
`outside' regions must be equal in magnitude but opposite
in sign.
Thus, the Hamiltonian cannot be positive definite.

One might imagine that there
is some way to define the ground state such that
the
{\em difference} between the quasilocal energy
of a given state and the quasilocal energy of
the ground state is positive definite.
However, by an argument similar to that
provided
above, one can show that there is no definition of the ground state
which would result in quasilocal energies which
are both additive and positive definite relative to the
ground.

These results are not so
disastrous as they seem.  It has been shown that
for asymptotically flat spacetimes, the quasilocal
energy (as evaluated at spacelike infinity
and relative to the Minkowski vacuum)
reduces to the Arnowitt--Deser--Misner
(ADM) energy~\cite{IKL}.
Also, in the local limit, the quasilocal energy
(relative to the Minkowski vacuum)
reduces to the three volume integral of the energy
density~\cite{Hayw/Wong1}.
Hence, there is no obvious conflict with the positivity
of the classical energy.

Before exploring matters further, let us review the Hamiltonian
formulation of general relativity for systems with timelike boundary
and the definition of quasilocal energy to which it gives rise.

\section{\label{S2}Derivation of Hamiltonian and quasilocal energy}

Let $\{g^{ab},{\cal M}\}$ be a spacetime
with topology
${\Sigma}_t\times\bar{{\sf I}}_t$, where the $\Sigma_t$
are closed, orientable spacelike
hypersurfaces and $\bar{\sf{I}}_t$
is a closed (timelike)
interval. Let ${\cal B}$ be the boundary of ${\cal M}$
and let $B_t$ be the (two dimensional) boundary of $\Sigma_t$.
By construction,
${\cal B}\equiv \Sigma_{t_0}\bigcup \Sigma_{t_1}
\bigcup B$
where $\Sigma_{t_0}$ and $\Sigma_{t_1}$
are initial and final spacelike
hypersurfaces and $B\equiv
B_t\times\bar{\sf I}_t$,
is the timelike boundary hypersurface.  See Figure 1(a).

Let $u^a$ be the
future pointing unit normal to $\Sigma_t$ and let
$n^a$ be the outward pointing normal to $B_t$ tangent
to $\Sigma_{t}$.
Similarly, let $\tilde{n}^a$ be the outward pointing normal
to $B$ and let $\tilde{u}^a$
be the future pointing normal to $B_{t}$
tangent to $B$.
Then, $h_{ab}=g_{ab}+u_a u_b$ is the
induced metric on $\Sigma_t$,
$\gamma_{ab}=g_{ab}-\tilde{n}_a \tilde{n}_b$ is
the induced metric on $B$, and
$\sigma_{ab}=\gamma_{ab}+\tilde{u}_a
\tilde{u}_b=h_{ab}-n_a n_b$
is the induced
metric on $B_t$.  See Figure 1(b).

The
action functional appropriate
to fixed intrinsic geometry on ${\cal B}$
is~\cite{York1972,Gibbons/Hawking,Hayw/Wong2,2bound}:
\begin{eqnarray}
I&=&{1\over16\pi}\int_{\cal
M}R\,{g}^{1/2}\,d^4x-{1\over8\pi}
\int_{\Sigma_{t_1}}K{h}^{1/2}\,d^3x
+{1\over8\pi}\int_{\Sigma_{t_0}}K{h}^{1/2}\,d^3x\nonumber\\
&&
\!\!\!\!\!\!\!\!\!\!\!\!+{1\over8\pi}
\int_{ B}\widetilde{K}
{\gamma}^{1/2}\,d^3x
+{1\over8\pi}\int_{{B}_{t_1}}\!\!\eta\,
{\sigma}^{1/2}\,d^2x
-{1\over8\pi}\int_{{B}_{t_0}}\!\!\eta\,
{\sigma}^{1/2}\,d^2x+I_{\rm m}+C.\label{newi}\end{eqnarray}
where
$K_{ab}\equiv h_{ac}\nabla^c u_b$
is the extrinsic curvature tensor on $\Sigma_{t}$
while $\widetilde{K}_{ab}\equiv
\gamma_{ab}\nabla^a \tilde{n}_b$
is the extrinsic curvature tensor on $B$ and
$\eta\equiv \sinh^{-1}{(u\cdot \tilde{n})}$
is the local boost parameter on $B_{t}$.
Also, $I_{\rm m}[A_a,P^a]$
is the action
associated with the
matter fields
and their conjugate momenta (which we label $A_a$ and $P^a$, respectively).

In (\ref{newi}), as with any action
functional, it is possible to add
a functional, $C$,
of the fields which are held fixed on the boundary
without affecting
the equations of motion.
The question of
whether there exists
a particularly useful
choice for $C$ in the gravitational context has received some
attention~\cite{Gibbons/Hawking,Brown/York}.
However, here we are interested only in variations of
the action between states
satisfying the same boundary conditions.
These variations are, of course, independent of $C$.
For simplicity of presentation
and without loss of generality,
we set $C=0$.

To obtain the Hamiltonian formulation of $I$,
choose a time flow vector field, $t^a$,
which satisfies
$t^a\nabla_a t=1$.
Also suppose $t^a$ is tangent to $B$ on $B_t$.
Define a lapse function, $N=-u_at^a$, and
a shift vector, $N^a=t^a-Nu^a$, on $\Sigma_t$.
Also define a lapse function,
$\widetilde{N}=-\tilde{u}_at^a$, and
a shift vector, $\widetilde{N}^a=
t^a-\widetilde{N}\tilde{u}^a$,
associated with the foliation of the intrinsic geometry of
$B$ into 2--surfaces,
$B_t$.
Further defining
a momentum field, $p^{ab}$,
conjugate to $h_{ab}$ in the usual fashion,
we obtain~\cite{Brown/York,Hayw/Wong2}\footnote{The treatment of
Ref.~\cite{Brown/York} does not include
the `kinetic' term,
${\eta\over8\pi}\,\,
\dot{\phantom{\sqrt{\sigma}}}\!\!\!\!\!\!\!\!\!\!\!\sqrt{\sigma}$,
in the boundary's contribution to the Lagrangian.
This
is because the authors assume that the timelike and spacelike
portions of the boundary intersect each other orthogonally
at $B_t$ and, hence, that $\eta=0$ there.},
\begin{eqnarray} I
&=&\int_{{\cal M}}{\cal L}\,d^3x\,dt
+\int_B{\cal L}_B\,d^2x\,dt\nonumber\\
&=&\int_{{\cal M}}\left(p^{ab}\dot{h}_{ab}
+P^a\dot{A}_a-N{\cal H}
-N^a{\cal H}_a-A_a{\cal G}^a\right)\,d^3x\,dt\nonumber\\
&& +\int_B\left({\eta\over8\pi}\,\,
\dot{\phantom{\sqrt{\sigma}}}\!\!\!\!\!\!\!\!\!\!\!
\sqrt{\sigma}
-\widetilde{N}\widetilde{{\cal H}}
-\widetilde{N}^a\widetilde{{\cal H}}_a-\widetilde{A}_a
\widetilde{{\cal G}}^a\right)\,d^2x\,dt
\label{newican}\end{eqnarray}
In the above, ${\cal H}$ and ${\cal H}_a$
correspond to the super--Hamiltonian and
super--momentum, respectively,
while the ${\cal G}^a$ are associated with
any constraints on the matter
fields. Also,  $\widetilde{{\cal H}}=-\left.
{\delta I\over\delta \widetilde{N}}\right|_{B}$
and
$\widetilde{{\cal H}}_a=
-\left.{\delta I\over\delta \widetilde{N}^a}
\right|_{B}$ are, respectively,
the `boundary super--Hamiltonian'
and `boundary super--momentum'~\cite{Hayw/Wong1,Hayw/Wong2},
while $\widetilde{{\cal G}}^a$ are
associated with any conserved
charge densities~\cite{Brown/York}.

Perform Legendre transformations\footnote{
Since the treatment of Ref.~\cite{Brown/York}
assumes $\eta=0$,
their definition of the Hamiltonian does not involve a Legendre
transformation with respect to the boundary `kinetic' term.  Nonetheless,
such a Legendre transformation is
necessary if the extremal Hamiltonian is not to be depend on the
gauge choice for the
spacelike slicing.}
on (\ref{newican}) to obtain the Hamiltonian,
\begin{eqnarray}
\!\!\!\!H\!\!\!&=&\!\!\!\!\int_{\Sigma_t}
\left(p^{ab}\dot{h}_{ab}
+P^a\dot{A}_a-{\cal L}\right)\,d^3x
+\int_{B_t}\left({\eta\over 8\pi}\,\,
\dot{\!\!\!\!\sqrt{\sigma}}
-{\cal L}_B\right)\,d^2x
\nonumber\\
&=&\!\!\!\!\int_{\Sigma_t}\left(N{\cal H}+N^a{\cal H}_a
+A_a {\cal G}^a\right)\,d^3x
+\int_{B_t}\left(\widetilde{N}
\widetilde{{\cal H}}+\widetilde{N}^a
\widetilde{{\cal H}}_a+\widetilde{A}_a
\widetilde{{\cal G}}^a
\right)\,d^2x.\label{Hammer}\end{eqnarray}
When the geometry is allowed to vary freely
everywhere on $\Sigma_t$,
we obtain the usual Hamilton's equations,
\begin{eqnarray}
\dot{h}_{ab}&=&
{\delta H\over \delta p^{ab}}
\nonumber\\
{\dot{p}^{ab}}&=&-{\delta H\over \delta h_{ab}}.
\label{hameqns}\end{eqnarray}
plus boundary counterparts,
\begin{eqnarray}
{\dot{\!\!\!\!\sqrt{\sigma}}\over 8\pi}&=&
\left.{\delta H\over \delta \eta}\right|_{B_t}
\nonumber\\
{\dot{\eta}\over 8\pi}&=&-\left.{\delta H\over \delta
\sqrt{\sigma}}\right|_{B_t}.
\label{boundhameqns}\end{eqnarray}

Define the quasilocal energy, $E$, as the
value of the Hamiltonian for the classical
spacetime configuration as measured by observers
for which $t^a=\tilde{u}^a$ on $B_t$ (e.g. as measured by
observers which travel orthogonal to $B_t$ on $B$).
This yields\footnote{
Expression (\ref{qlee1}) for the quasilocal energy
was obtained in Ref's~\cite{Hayw/Wong1,Hayw/Wong2}.
This expression also agrees with
the expression for the quasilocal energy obtained
by Brown and York~\cite{Brown/York}
when
$\Sigma_{t}$ and $B$ are orthogonal
on $B_{t}$ (i.e. $n^a=\tilde{n}^a$)
but does {\em not} agree with
their expression otherwise.},
\begin{equation}
E=\int_{B_{t}}
\left.\widetilde{{\cal H}}\right|_{\rm cl}\,d^2x=-{1\over8\pi}\int_{B_{t}}
\sigma_a^b\nabla^a \tilde{n}_b
\sqrt{\sigma}\,d^2x.\label{qlee1}\end{equation}

It is also possible to define
the quasilocal energy and, more generally, a
quasilocal momentum field in terms
of a boundary surface stress tensor~\cite{Brown/York}.
This formulation employs a natural extension of
classical Hamilton--Jacobi theory
and is valuable to review because
we shall find that the quasilocal dominant
energy condition can be expressed in terms of
variations of the boundary stress tensor.

Let $S\equiv \left. I\right|_{\rm extremum}$
be the extremal action for a system subject to
constrained geometry along $B$.
Define a boundary surface stress tensor,
$\tau^{ab}$, in terms of the variations of $S$
induced by varying the constrained boundary
geometry~\cite{Brown/York},
\begin{equation}
\tau^{ab}\equiv \left.{2\over\sqrt{\gamma}}
{\delta S\over \delta \gamma_{ab}}\right|_B.
\nonumber\end{equation}

Choose a time flow vector such that
$\widetilde{N}$ is constant over $B_{t}$.
Let $T$ be the lapse of proper time as measured from
$B_{t}$ to a neighboring slice $B_{t+\Delta t}$
by observers traveling orthogonal to $B_{t}$.
Then, uniformly varying $T$ over $B_{t}$ and
taking the limit $\Delta t\to 0$, define the quasilocal
energy by,
\begin{equation}
E\equiv-\left.{\delta S\over\delta T}\right|_{B_{t}}
=\int_{B_{t}}
\left.\widetilde{{\cal H}}\right|_{\rm cl}\,d^2x=
\int_{B_{t}}
\tilde{u}^a\tilde{u}^b\tau_{ab}\sqrt{\sigma}\,d^2x.
\label{qlee}\end{equation}
Thus, the quasilocal energy is the
negative of the variation
of the extremal action induced by uniformly varying the
interval
of proper time as measured by observers traveling
orthogonal to $B_t$ on $B$.  This definition
naturally extends the definition of energy
that arises in classical Hamilton--Jacobi theory.

More generally, define a
quasilocal momentum field tangent
to $B$ as follows.
Let $e^a_{(i)}$ be a triad tangent
to $B$ (for $i=0,1,2$) such that
$e^a_{(0)}\equiv \tilde{u}^a$ and $e^a_{(1)}$
and $e^a_{(2)}$ are
tangent to $B_t$.  We can then express $t^a$ as
$t^a=t^ie^a_{(i)}$.
Choosing $t^a$ such that $t^i$ are constant over $B_t$,
define components of a quasilocal momentum field
tangent to $B$ in the limit $\Delta t\to 0$ by,
\begin{eqnarray}{\cal P}_i&=&
\left.{\delta S\over\delta T^i}\right|_{B_{t}}
\nonumber\\
&=&
-\int_{B_t}e^a_{(i)}\tilde{u}^b\tau^{ab}
\sqrt{\sigma}\,d^2x,\end{eqnarray}
where $T^i=t^i\Delta t$.

\section{\label{S3}Quasilocal Energy Conditions}
In this section, we establish
that the local
dominant energy condition
gives rise to a quasilocal counterpart.
Let us begin by deriving an expression
for the variation of
the quasilocal momentum field
induced when we perturb a vacuum state
to obtain a new state satisfying the same boundary conditions.

Consider a manifold ${\cal M}$ of the type
discussed in the previous section.
Constrain the geometry on its timelike
boundary $B$.
Let $\{{g}^{ab}\}$ be the class of all smooth metrics
which
cover ${\cal M}$ and have the appropriate
geometry on $B$.  Let $\{g^{ab}_{\rm vac}\}$ be the
sub--class of these metrics composed of all solutions
which are vacuum states of the matter field
(i.e. solutions to Einstein's equations with $T^{ab}=0$).
Depending on the choice of the fixed geometry on
$B$, $\{g^{ab}_{\rm vac}\}$ may be a finite set, an infinite
set, or the empty set.

Take variations of the
Hamiltonian (\ref{Hammer}) around $g^{ab}_{\rm vac}$
and impose Hamilton's
equations (\ref{hameqns}) and their boundary
counterparts (\ref{boundhameqns}) to obtain,
\begin{eqnarray}
\!\!\!\!\int_{B_{t}}\!\!\left(\widetilde{N}\delta{\widetilde{H}}
\!\!+\!\!\widetilde{N}^a\delta \widetilde{H}_a\right)
d^2x\!\!&=&\!\!\int_{\Sigma_{t}}\!\!
\left\{\dot{h}_{ab}\delta p^{ab}
\!\!-\!\!\dot{p}^{ab}\delta h_{ab}\!\!-\!\!N\delta
{\cal H}_{\rm (m)}-N^a\delta
{\cal H}_{a {\rm (m)}}
\right\}d^3x\nonumber\\
&&
+{1\over 8\pi}
\int_{B_{t}}\left\{\,\,\dot{\!\!\!\!
\sqrt{\sigma}}\,\delta \eta-\dot{\eta}
\,\delta\sqrt{\sigma}\right\}\,d^2x\label{varham}\end{eqnarray}
where $N{\cal H}_{\rm (m)}+
N^a{\cal H}_{a {\rm (m)}}=
u^at^bT_{ab}\sqrt{h}$ is the matter's
contribution to the Hamiltonian.

Equation (\ref{varham}) is a general expression for the variation
of the Hamiltonian around a vacuum solution.
It can also be derived by taking variations of the
microcanonical action~\cite{microcanonical}
around $g^{ab}_{\rm vac}$ and taking the limit
$\Delta t\equiv t_1-t_0\to 0$.

Now let us further refine the scope of our
attention to variations
of the Hamiltonian around an {\em extremum of its vacuum
values}.
Define
$\{\bar{g}^{ab}\}$ to be the set
of solutions which extremize
$H$ over $\{g^{ab}_{\rm vac}\}$.
We will refer to any such states
as `ground states' compatible
with
the fixed geometry on $B$.
{}From (\ref{varham}) we obtain
that the variation of $H$ around a ground state
is given by,
\begin{equation}
\int_{B_{t}}\left(\widetilde{N}\delta{\widetilde{H}}
+\widetilde{N}^a\delta \widetilde{H}_a\right)
d^2x
=\int_{\Sigma_{t}}
u^at^b\delta
T_{ab}\,\sqrt{h}\,d^3x
.\label{varham22}\end{equation}
Note that $\tilde{u}_a{t}_b\tau^{ab}
\sqrt{\sigma}=\widetilde{N}\widetilde{{\cal H}}+
\widetilde{N}^a\widetilde{{\cal H}}_a$, so equation (\ref{varham22})
can also be expressed in the form,
\begin{equation}
\int_{B_{t}}\tilde{u}^at^b\delta \tau_{ab}\,\sqrt{\sigma}d^2x=\int_{\Sigma_{t}}
u^at^b\delta
T_{ab}\,\sqrt{h}\,d^3x.\label{varham222}\end{equation}

Finally, let us restrict the choice of
the time flow vector on the boundary so that
the quasilocal momentum field is well defined
in the sense described in Section \ref{S2}
(this is equivalent to restricting the gauge
choice so that the boundary lapse and shift are
constant over $B_t$).
We obtain from (\ref{varham222}),
\begin{equation} -t^i\,\delta{\cal P}_i=\int_{\Sigma_{t}}
\left\{u^at^b\delta
T_{ab}\,\sqrt{h}\right\}\,d^3x,
\label{interm2}\end{equation}
where the variation is conducted around
a ground state compatible with a given fixed geometry
on $B$.
When the local dominant energy condition is satisfied
everywhere on $\Sigma_t$, we have $u^at^b\delta
T_{ab}\ge0$ (since variations are
performed around a $T^{ab}=0$ solution).
In this case, (\ref{interm2}) implies,
\begin{equation}
t^i\delta {\cal P}_i=t^i({\cal P}^i-\bar{{\cal P}}^i)\le
0.\label{qlec3}\end{equation}

We interpret equation (\ref{qlec3})
as a quasilocal dominant energy condition.
It requires that
perturbations from a ground state, $\bar{g}^{ab}$,
which preserve the fixed geometry of $B$
cannot give rise to a spacelike
variation of the quasilocal momentum vector ${\cal P}^i$.

\section{\label{S4}Discussion}
Let us examine more closely the significance of
the quasilocal energy condition (\ref{qlec3}).

First, it is important to stress that equation (\ref{qlec3})
in general only holds for variations conducted around a
ground state (i.e. a vacuum state
which extremizes the Hamiltonian
over $\{g^{ab}_{\rm vac}\}$).
In fact,
it is easy to see that equation (\ref{qlec3}) can be violated
when variations are conducted around a vacuum solution
which is not extremal.
Perturbations from one vacuum solution to another
would generate linear variations in the Hamiltonian and, hence, must
in some cases
result in a decrease of the quasilocal energy.
[An explicit example which demonstrates this is given in the next section.]

Second, note that for a given fixed geometry along $B$, there is no
guarantee that a ground state will exist.  Nor is there
any guarantee that when a ground state
exists it is unique.  In principle, there
might even even be an infinite set of
ground states compatible with a given boundary geometry.

Third, note that equation (\ref{qlec3})
holds only for {\em local} perturbations around
a ground state.
It
does not necessarily imply that a ground state
is a global
minimum energy
state,
nor does it necessarily imply the existence of
a finite least energy state (see example of next section).
In some cases,
a given fixed geometry on $B$ might give
rise to an infinite class of vacuum solutions
for which the energy could be arbitrarily negative.

Fourth, note that the quasilocal dominant energy condition
yields a quasi\-local weak energy condition as a consequence.
Specifically, if we choose a time flow vector parallel to
$\tilde{u}^a$, equation (\ref{qlec3}) yields
\begin{equation}
\delta E=E-\bar{E}\ge0
\label{qlec}\end{equation}
where $E$ is the quasilocal
energy for the system with matter distribution
and $\bar{E}$ is the quasilocal energy for the ground state.

\section{\label{S5} Quasilocal energy condition
for spacetimes
with Schwarzschild geometry.}

An explicit example will help clarify the discussion of
the previous section.

Let $\{g^{ab}, {\cal M}\}$ be a spacetime
with Schwarzschild geometry,
\begin{equation}
ds^2=-\left(1-2M/r\right)\,d\hat{t}^2
+\left(1-2M/r\right)^{-1}\,dr^2+r^2\,d\Omega^2.
\label{schwarz}
\end{equation}
Further suppose the spacetime extends between
timelike tubes at $r=r_0$ and $r=r_1$.

To evaluate the quasilocal energy
as measured by observers
on a given boundary 2--surface, we must
choose a time flow vector such that the
lapse and shift are constant over the boundary 2--surface.
[Note that the lapse is not constant over $B_{\hat{t}}$
for $\hat{t}^a=\left({\partial \over\partial \hat{t}}\right)^a$.]
Let us take $t^a\equiv \left({\partial \over\partial
t}\right)^a$ where
\begin{equation}t-t_1=\left(1-2M/r\right)^{1/2}\hat{t}
\end{equation}
and evaluate the quasilocal energy over the boundary
2-surface at $t=t_1$.
In this frame, the line element (\ref{schwarz}) becomes
\begin{equation}
ds^2=-dt^2+{2M(t-t_1)\over r^2 (1-2M/r)}\,dr\,dt
+\left[{1\over 1-2M/r}-{M^2(t-t_1)^2\over r^4(1-2M/r)^2}\right]
\,dr^2+r^2\,d\Omega^2.\label{newschwarz}\end{equation}

Let us fix the geometry on $B$ so that its intrinsic
line element is
\begin{equation}
ds^2=-dt^2+r^2\,d\Omega^2\label{intrin}\end{equation}
for $r=r_0$ and $r=r_1$.
{}From (\ref{newschwarz}) it is clear that
the Schwarzschild solutions with $2M <r_0$
provide isometric embeddings of the boundary geometry.
Hence, $\{g^{ab}_{\rm vac}\}$ consists of all Schwarzschild
solutions with $2M<r_0<r_1$.

Evaluating expression (\ref{varham})
for variations in the Hamiltonian
over this class of vacuum solutions
yields,
\begin{equation}
\delta E=\left.(1-2M/r)^{-1/2}
\right|^{r_1}_{r_0}\,\delta M.\label{exam}\end{equation}

Equation (\ref{exam}) implies
that flat spacetime is the unique ground state compatible
with these boundary conditions (i.e. $\{\bar{g}^{ab}\}$
consists of a unique element which is the flat metric).
Thus, when the geometry on the timelike boundary
is held fixed with line element given by (\ref{intrin}),
the quasilocal dominant energy condition
implies
$\delta E\ge0$ for linear perturbations
from
flat spacetime so long as the local dominant energy
condition is everywhere satisfied.

Note, however, that the quasilocal energy condition
need not hold for variations around vacuum solutions
which do not extremize the Hamiltonian.
In particular, when both $M$ and $\delta M$ are greater
than zero, equation (\ref{exam})
yields $\delta E<0$.

Also note that even though
$\delta E\ge0$ for linear perturbations
from
flat spacetime, the quasilocal energy associated with flat
spacetime is neither a minimum nor a lower bound
to the energies accessible with the boundary
geometry (\ref{intrin}).
In fact, if we allow for vacuum solutions with $M<0$,
the quasilocal energy is not bounded from below: $E\to -\infty$
as $M\to -\infty$.
On the other hand, if we confine attention to solutions with
$M\ge0$, the quasilocal energy achieves a lower bound
for the vacuum solution with $M=r_0/2$.

\section{\label{S6} Conclusions}
The quasilocal energy (i.e. the classical
value of the Hamiltonian) is not positive definite.
Nor is there any way to define a
ground state such that
the quasilocal energy
of a given state relative to the
ground is both positive definite and additive.
In fact, it may be (as
in the example of Section (\ref{S5})) that
the set of quasilocal energies accessible
to a system with a given boundary
geometry is not bounded from below.

Despite all this, we have demonstrated
that a quasilocal dominant energy condition
holds for linear perturbations around
a ground state so long as the local dominant
energy condition is everywhere satisfied.
The quasilocal dominant energy condition
also implies a quasilocal weak energy condition.

{\bf Acknowledgments}: I am very much indebted to
both Stephen Anco and Ken Wong.
Much research which laid the foundations
for this paper was done in collaboration with them.
In particular, Stephen Anco contributed much of
the formalism used here to derive
the quasilocal energy and
momentum.  I am also very grateful to
Warren Anderson, Andrei Barvinsky, Werner Israel,
and, in particular, Bill Unruh and
Don Page for many useful discussions.
This research was supported by
the National Science and Engineering
Research Council of Canada
and the Canadian Institute for Theoretical
Astrophysics.

\eject
\centerline{\bf List of Figure Captions}
\vspace{.5cm}
\noindent
{\bf Figure 1}:  (a) A spacetime
extending from spacelike hypersurface
$\Sigma_{t_0}$ to spacelike
hypersurface $\Sigma_{t_1}$ and out to
timelike boundary $B$.  The
timelike and spacelike portions of the boundary
intersect at $B_{t_0}$ and $B_{t_1}$.
(b) The same spacetime
with the various unit normals
and boundary metrics indicated.

\baselineskip=4.mm
\begin{thebibliography}{99}
\bibitem{Brown/York}J. D. Brown and J. W. York. Jr.,
Phys. Rev. D {\bf 47} 1407 (1993).

\bibitem{Euclidean}  G. Hayward, Phys. Rev. D {\bf 41}
3248 (1990).

\bibitem{IKL} J. Katz, D.
Lynden--Bell and W. Israel,
Class. Quantum Grav. {\bf 5}, 971 (1988).

\bibitem{Hayw/Wong1}
G.~Hayward and Ken Wong, Phys. Rev. D {\bf 46}, 620 (1992).

\bibitem{York1972}
J.~W. York, Jr., Phys. Rev. Lett. {\bf 28}, 1082 (1972).

\bibitem{Gibbons/Hawking} G. W.
Gibbons and S. W. Hawking, Phys. Rev. D {\bf 15}, 2752 (1977).

\bibitem{Hayw/Wong2}
G.~Hayward and Ken Wong, Phys. Rev. D {\bf 47}, 4778 (1993).

\bibitem{2bound} G. Hayward, Phys. Rev. D {\bf 47}, 3275 (1993).

\bibitem{microcanonical} J. D. Brown and J. W. York. Jr.,
Phys. Rev. D {\bf 47} 1420 (1993).

\end{thebibliography}
\end{document}